\documentclass[aps,prl,preprint,showpacs,groupedaddress]{revtex4}
\usepackage {amssymb}
\usepackage {amsmath}
\usepackage {graphicx}
\usepackage {longtable}

\begin{document}
\title{The dissolution of the vacancy gas and grain boundary diffusion
in crystalline solids}

\author{Fedor V.Prigara}
\affiliation{Institute of Microelectronics and Informatics,
Russian Academy of Sciences,\\ 21 Universitetskaya, Yaroslavl
150007, Russia} \email{fvprigara@rambler.ru}

\date{\today}

\begin{abstract}

Based on the formula for the number density of vacancies in a solid under
the stress or tension, the model of grain boundary diffusion in crystalline
solids is developed. We obtain the activation energy of grain boundary
diffusion (dependent on the surface tension or the energy of the grain
boundary) and also the distributions of vacancies and the diffusing species
in the vicinity of the grain boundary.

\end{abstract}


\maketitle

Recently, it was shown that sufficiently high pressures as well as
mechanical stresses applied to a crystalline solid lead to the decrease in
the energy of the vacancy formation and create, therefore, an additional
amount of vacancies in the solid [1]. The last effect enhances
self-diffusion in the crystal which is normally vacancy-mediated, at least
in simple metals. Since large mechanical stresses are normally present in
grain boundaries, these new results can elucidate the mechanisms of grain
boundary diffusion which have remained so far unclear [2].

According to the thermodynamic equation [3]

\begin{equation}
\label{eq1}
dE = TdS - pdV,
\end{equation}

\noindent
where \textit{E} is the energy, \textit{T} is the temperature, \textit{S} is
the entropy, \textit{p} is the pressure, and \textit{V} is the volume of a
solid, the energy of a solid increases with pressure, so the pressure acts
as the energy factor similarly to the temperature. Therefore, the number of
vacancies in a solid increases both with temperature and with pressure.

The thermodynamic consideration based on the Clausius- Clapeyron equation
gives the number density \textit{n} of vacancies in a solid in the form [1]

\begin{equation}
\label{eq2}
n = \left( {P_{0} /T} \right)exp\left( { - E_{v} /T} \right) = \left( {n_{0}
T_{0} /T} \right)exp\left( { - E_{v} /T} \right),
\end{equation}

\noindent where $E_{v} $ is the energy of the vacancy formation,
$P_{0} = n_{0} T_{0} $ is a constant, $T_{0} $ can be put equal to
the melting temperature of the solid at ambient pressure, and the
constant $n_{0} $ has an order of magnitude of the number density
of atoms in the solid. Here the Boltzmann constant $k_{B} $ is
included in the definition of the temperature \textit{T}.

The formula (\ref{eq2}) describes the thermal expansion of the solid. It should be
taken into account that the dissolution of the vacancy gas in a solid causes
the deformation of the crystalline lattice and changes the lattice
parameters.

The energy of the vacancy formation $E_{v} $ depends linearly on the
pressure \textit{P} (in the region of high pressures) as given by the
formula

\begin{equation}
\label{eq3}
E_{v} = E_{0} - \alpha P/n_{0} ,
\end{equation}

\noindent
where $\alpha $ is a dimensionless constant, $\alpha \approx 18$ for
sufficiently high pressures. On the atomic scale, the pressure dependence of
the energy of the vacancy formation in the equation (\ref{eq3}) is produced by the
strong atomic relaxation in a crystalline solid under high pressure.

With increasing pressure, the number density of vacancies in a solid
increases, according to the relation

\begin{equation}
\label{eq4}
n = \left( {n_{0} T_{0} /T} \right)exp\left( { - \left( {E_{0} - \alpha
P/n_{0}}  \right)/T} \right),
\end{equation}

\noindent
and, finally, the vacancies can condense, forming their own sub-lattice.
Such is the explanation of the appearance of composite incommensurate
structures in metals and some other elemental solids under high pressure
[4-7].

Further increase of the number density of vacancies in a solid with
increasing pressure leads to the melting of the solid under sufficiently
high pressure (and fixed temperature). Such effect has been observed in
sodium [6]. In general, such behavior is universal for solids, though the
corresponding melting pressure is typically much larger than those for
sodium.

We assume that the melting of the crystalline solid occurs when the critical
number density $n_{c} $ of vacancies is achieved. In view of the equation
(\ref{eq2}), it means that the ratio of the energy of the vacancy formation $E_{v} $
to the melting temperature $T_{m} $ of the solid is approximately constant,

\begin{equation}
\label{eq5}
E_{v} /T_{m} \approx \alpha .
\end{equation}

The value of the constant $\alpha $ in the last relation can be determined
from the empirical relation between the activation energy of self diffusion
(which is approximately equal to the energy of vacancy formation) and the
melting temperature of a solid [8]:

\begin{equation}
\label{eq6}
E_{0} \approx 18T_{m} ,
\end{equation}

\noindent
so that $\alpha \approx 18$.

Substituting the expression (\ref{eq3}) in the relation (\ref{eq5}), we obtain

\begin{equation}
\label{eq7}
\left( {E_{0} - \alpha P/n_{0}}  \right)/T_{m} \approx \alpha .
\end{equation}

The last equation gives the melting curve of the crystalline solid in the
region of high pressures in the form

\begin{equation}
\label{eq8} T + P/n_{0} \approx E_{0} /\alpha \approx T_{0} ,
\end{equation}

\noindent
where $T_{0} $ is the melting temperature of the solid at ambient pressure.

The constant $n_{0} $ can be determined from the relation between the
tensile strength $\sigma _{s} $ and the melting temperature $T_{m} $ of a
solid [1]

\begin{equation}
\label{eq9}
n_{0} \cong \sigma _{s} /T_{m} .
\end{equation}

The numerical value of this constant is $n_{0} \approx 1.1 \times
10^{22}cm^{ - 3}$ [1].

Replacing in the relation (\ref{eq4}) the pressure \textit{P} by
the absolute value of the stress or tension $\sigma = F/S$,
applied to a solid, where \textit{F }is the applied force and
\textit{S} is the cross-section area of the solid in the plane
perpendicular to the direction of the applied force, we can
estimate the mean number density of vacancies in the solid under
the stress or tension:

\begin{equation}
\label{eq10}
\langle n\rangle \cong \left( {n_{0} T_{0} /T} \right)exp\left( { - \left(
{E_{0} - \alpha \sigma /n_{0}}  \right)/T} \right).
\end{equation}

The dissolution of the vacancy gas in a solid under the stress or tension is
responsible for the low values of the elastic limit and the tensile strength
of solids as compared with theoretical estimations not taking into account
this process [9].

As indicated above, large mechanical stresses are normally present
in grain boundaries. The absolute value $\sigma _{b} $ of the
mechanical stress in the close vicinity of a grain boundary is
given by the formula

\begin{equation}
\label{eq11} \sigma _{b} \cong \gamma _{b} /r_{0} ,
\end{equation}

\noindent
where $\gamma _{b} $ is the energy of the grain boundary and $r_{0} $ is the
radius of the atomic relaxation region (around a vacancy) which will be
estimated below.

According to the relation (\ref{eq10}), the energy of the vacancy
formation in the close vicinity of the grain boundary is given by
the formula

\begin{equation}
\label{eq12}
E_{b} = E_{0} - \alpha \gamma _{b} /\left( {n_{0} r_{0}}  \right).
\end{equation}

For the small values of misorientation angle $\theta \leqslant 10
- 15^{}$ degrees, the energy of the dislocation structure
contributes to the energy of the grain boundary [10]. However, for
larger misorientation angles, the energy of the grain boundary is
approximately constant and is determined by the surface tension
$\gamma $ of the solid, $\gamma _{b} \cong \gamma $.

Due to the Einstein relation between the mobility of an atom, $\mu
= v/F$, where \textit{v} is the velocity of the atom and
\textit{F} is the force acting on the atom, and the diffusion
coefficient \textit{D} [8]:

\begin{equation}
\label{eq13} \mu = v/F = D/T,
\end{equation}

\noindent the speed of grain boundary motion \textit{v} is
proportional to the diffusion coefficient $D_{ \bot}  $ for
self-diffusion in the direction perpendicular to the plane of a
grain boundary. Therefore, the activation energy \textit{E} of
grain boundary motion is equal to the activation energy $E_{ \bot}
$ of self-diffusion across the grain boundary. The last activation
energy is equal to the activation energy $E_{b} $ of grain
boundary self-diffusion in the case of high-angle grain
boundaries, and is approximately equal to the activation energy
$E_{0} $ of bulk self-diffusion for low-angle grain boundaries.
Thus, there is a step of the activation energy for grain boundary
motion at some critical value $\theta _{c} $ of the misorientation
angle ($\theta _{c} = 10 - 15^{ \circ} $, as indicated above).
Such a step of the activation energy for grain boundary motion has
been observed experimentally in high-purity aluminium, the
critical value of the misorientation angle being in this case
$\theta _{c} = 13.6^{ \circ} $ [11].

The driving force for grain boundary motion is provided by the
distribution of mechanical stresses in a crystalline solid [12].

Assuming that the free surface of a crystalline solid is formed by the plane
of vacancies, we can estimate the surface tension of the solid as follows

\begin{equation}
\label{eq14} \gamma \cong \beta n_{0} E_{0} a_{0},
\end{equation}

\noindent where $a_{0}=n_{0}^{-1/3} \cong 0.45nm $ has an order of
magnitude of the lattice spacing $a $, and $\beta $ is a
dimensionless constant which has an order of unity. For hard
metals such as Al, Zr, Nb, Fe, Pt, $\beta \cong 0.8 $. In the case
of mild metals, $\beta $ is normally smaller, e.g. for Rb and Sr,
$\beta \cong 1/4 $.

Substituting the estimation (\ref{eq14}) for the energy of the
grain boundary $\gamma _{b} \cong \gamma $ in the equation
(\ref{eq12}), we find

\begin{equation}
\label{eq15} E_{b} \approx E_{0} \left( {1 - \beta \alpha
a_{0}/r_{0}} \right).
\end{equation}

Due to the atomic relaxation and thermal motion of atoms, the
migration barriers are small [2,13], and the activation energy of
self-diffusion is approximately equal to the energy of the vacancy
formation. The analysis of experimental data on the activation
energy of grain boundary self-diffusion gives an empirical
relation [14]

\begin{equation}
\label{eq16} E_{b} \approx 9T_{m} \approx E_{0} /2.
\end{equation}

From equations (\ref{eq15}) and (\ref{eq16}), we find the
estimation of the radius of the atomic relaxation region,

\begin{equation}
\label{eq17} r_{0} \approx 2\beta \alpha a_{0} \cong \alpha a_{0},
\end{equation}

\noindent since $\beta $ has an order of unity. The radius of the
atomic relaxation region has an order of $r_{0} \cong 18n_{0}^{
-1/ 3} \approx 8nm$. This value is comparable with the diameters
of tracks produced by high energy ions in metals [15-17]. The
grain boundary diffusion width $\delta $ [14] is smaller than the
radius of the atomic relaxation region due to the non-uniform
distribution of vacancies inside the atomic relaxation region in
the grain boundary.

If we assume that the mechanical stress $\sigma $ decreases linearly with
the distance \textit{x} from the plane of the grain boundary,

\begin{equation}
\label{eq18} \sigma = \sigma _{0} \left( {1 - kx} \right),
\end{equation}

\noindent where $\sigma _{0} $ is the stress at the boundary of
the atomic relaxation region with the width $r_{0} $ in the grain
boundary (this value is smaller than $\sigma _{b} \cong \gamma
/r_{0}\cong\left(1/2\right)n_{0}T_{m} $ and has an order of
magnitude $\sigma _{0} \cong \left(1/2\right)n_{0} T$), then the
equation (\ref{eq10}) gives the distribution of vacancies in the
vicinity of the grain boundary in the form

\begin{equation}
\label{eq19} n \cong \left( {n_{0} T_{0} /T} \right)exp\left( { -
\left( {E_{0} - \alpha \sigma _{0} \left( {1 - kx} \right)/n_{0}}
\right)/T} \right) = n_{b} exp\left( { - \alpha \sigma _{0}
kx/\left( {n_{0} T} \right)} \right),
\end{equation}

\noindent
where $n_{b} $ is the number density of vacancies at the boundary of the
atomic relaxation region.

Due to the trapping by vacancies [18], the distribution of the
concentration \textit{c} of the diffusing species in the vicinity
of the grain boundary follows the same law:

\begin{equation}
\label{eq20} c \cong c_{b} exp\left( { - x/l} \right),
\end{equation}

\noindent
where $c_{b} $ is the concentration of the diffusing species at the boundary
of the relaxation region, and the scale \textit{l} is given by the formula

\begin{equation}
\label{eq21} l = n_{0} T/\left( {\alpha \sigma _{0} k} \right).
\end{equation}

Here \textit{k} has an order of magnitude of $1/d$, \textit{d}
being the size of the grain, so that $l \cong d/\alpha $. The
penetration profiles described by the equation (\ref{eq20}) have
been indeed observed experimentally in the case of grain boundary
diffusion in metals [8, 18], the measured penetration depth $ l$
having an order of a few micrometers [8].

To summerize, we obtained the dependence of the activation energy of grain
boundary self-diffusion on the energy of the grain boundary, the estimation
of the surface tension of a solid and of the energy of the grain boundary,
and the width of the atomic relaxation region in the grain boundary (or the
radius of the atomic relaxation region around a vacancy). We obtained
further the distributions of vacancies and the diffusing species in the
vicinity of the grain boundary. The obtained radius of the atomic relaxation
region is consistent with the diameters of tracks produced by high energy
ions in metals.

\begin{center}
---------------------------------------------------------------
\end{center}

[1] F.V.Prigara, E-print archives, cond-mat/0701148.

[2] A.Suzuki and Y.Mishin, J. Mater. Sci. \textbf{40}, 3155 (2005).

[3] S.-K.Ma, \textit{Statistical Mechanics} (World Scientific, Philadelphia,
1985).

[4] R.J.Nelmes, D.R.Allan, M.I.McMahon, and S.A.Belmonte, Phys. Rev. Lett.
\textbf{83}, 4081 (1999).

[5] M.I.McMahon, S.Rekhi, and R.J.Nelmes, Phys. Rev. Lett. \textbf{87},
055501 (2001).

[6] O.Degtyareva, E.Gregoryanz, M.Somayazulu, H.K.Mao, and R.J.Hemley, Phys.
Rev. B \textbf{71}, 214104 (2005).

[7] V.F.Degtyareva, Usp. Fiz. Nauk \textbf{176}, 383 (2006)
[Physics- Uspekhi \textbf{49}, 369 (2006)].

[8] B.S.Bokstein, S.Z.Bokstein, and A.A.Zhukhovitsky, \textit{Thermodynamics
and Kinetics of Diffusion in Solids} (Metallurgiya Publishers, Moscow,
1974).

[9] G.I.Epifanov, \textit{Solid State Physics} (Higher School Publishers,
Moscow, 1977).

[10] A.A.Smirnov, \textit{Kinetic Theory of Metals} (Nauka, Moscow, 1966).

[11] M.Winning, G.Gottstein, and L.S.Shvindlerman, Acta Mater.
\textbf{49}, 211 (2001).

[12] K.J.Draheim and G.Gottstein, in \textit{APS Annual March
Meeting, 17- 21 March 1997}, Abstract D41.87.

[13] B.P.Uberuaga, G.Henkelman, H.Jonsson, S.T.Dunham, W.Windl,
and R.Stumpf, Phys. Stat. Sol. B \textbf{233}, 24 (2002).

[14] I.Kaur and W.Gust, \textit{Fundamentals of Grain and
Interphase Boundary Diffusion} (Ziegler Press, Stuttgart, 1989).

[15] F.F.Komarov, Usp. Fiz. Nauk \textbf{173}, 1287 (2003)
[Physics- Uspekhi \textbf{46}, 1253 (2003)].

[16] F.V.Prigara, E-print archives, cond-mat/0406222.

[17] M.Toulemonde, C.Trautmann, E.Balanzat, K.Hjort, and
A.Weidinger, Nucl. Instrum. Meth. B\textbf{ 217}, 7 (2004).

[18] W.P.Ellis and N.H.Nachtrieb, J. Appl. Phys. \textbf{40}, 472
(1969).

\end{document}